\documentstyle[12pt]{article}
\def\la{\mathrel{\mathpalette\fun <}}
\def\ga{\mathrel{\mathpalette\fun >}}
\def\fun#1#2{\lower3.6pt\vbox{\baselineskip0pt\lineskip.9pt
  \ialign{$\mathsurround=0pt#1\hfil##\hfil$\crcr#2\crcr\sim\crcr}}}
\begin{document}
\pagestyle{empty}
\begin{center}
\vspace*{2.5cm}
\rightline{FERMILAB-Pub-95/274-A}
\rightline{Submitted to {\it Physical Review Letters}}
\vspace{1.5cm}
{\Large \bf WEAK INTERACTIONS IN SUPERNOVA CORES}\\
\bigskip
{\Large \bf AND SATURATION OF}\\
\bigskip
{\Large \bf  NUCLEON SPIN FLUCTUATIONS}

\vspace{.3in}

G\"unter~Sigl$^{a,b}$\\

\vspace{0.2in}

{\it $^a$Department of Astronomy \& Astrophysics\\
Enrico Fermi Institute, The University of Chicago, Chicago,
IL~~60637-1433}\\

\vspace{0.1in}

{\it $^b$NASA/Fermilab Astrophysics Center\\
Fermi National Accelerator Laboratory, Batavia, IL~~60510-0500}\\

\end{center}

\vspace{0.3in}

\centerline{\bf ABSTRACT}
\medskip
Extrapolation of perturbative nucleon spin fluctuation
rates seems to suggest a strong suppression of
weak interactions
in supernova cores. We derive a new sum rule for
the dynamical spin-density structure function which relates
the spin fluctuation rate to the
average nuclear interaction energy. For a bremsstrahlung like
structure function profile we show that instead of
strongly decreasing, the neutrino scattering cross section is
roughly density independent and axion
emission rates increase somewhat slower than the lowest order
emissivities towards the center of a hot supernova core.

\newpage
\pagestyle{plain}
\setcounter{page}{1}

\section{Introduction}
The cooling history of a newly born neutron star in the center
of a supernova (SN) is mainly determined by neutrino diffusion.
Numerical simulations employing the lowest order neutrino
interaction rates calculated within the Glashow-Salam-Weinberg
theory predict a cooling time scale which agrees remarkably well
with the neutrino signal observed from SN 1987 A~\cite{Schramm}.
The emission of novel weakly interacting particles like
axions~\cite{R3}
could change the cooling time scale substantially which in turn
allows to derive constraints on the properties of such
particles~\cite{Burrows}.

Within linear response theory weak interaction rates
with a medium of nonrelativistic nucleons are determined, apart
from the weak phase space, by only two dynamical structure
functions, one for the density and one for the nucleon
spin-density~\cite{Iwamoto1,R2}. Some work has
been devoted to their calculation but either the Landau theory
of quasiparticles was applied assuming
a ``cold'' nuclear medium~\cite{Iwamoto1,cold} or the authors
focused on quasielastic scattering studying static structure
functions~\cite{Iwamoto1,Sawyer,hot}.
Interactions of neutrinos and axions with a nonrelativistic
nuclear medium are mainly governed by the local nucleon spin-density and its
fluctuations. To lowest nontrivial order in the spin dependent
nucleon-nucleon interactions causing these fluctuations, the
relevant weak processes are of the nucleon bremsstrahlung type.
Due to the Landau
Pomeranchuk Migdal (LPM) effect~\cite{LPM} which accounts for multiple
nucleon scattering the inelasticity of these processes depends
on the nucleon spin flip rate. In addition, once this rate
becomes
considerably larger than the medium temperature $T$, the total
weak interaction rates tend to be suppressed~\cite{R1}.
Since perturbative estimates for the nucleon spin flip rate
can be as high as $\simeq50T$ around nuclear densities, this
could have profound implications for SN core
physics~\cite{R2,R1,Keil,Janka}.
By dramatically reducing the predicted SN
cooling time scale it would spoil the agreement between theory
and the observed neutrino pulse from SN 1987 A~\cite{Keil}. On
these phenomenological grounds it has been suggested that
axial-vector neutrino scattering cross sections might be roughly density
independent~\cite{Janka} instead of being suppressed at high
densities by the LPM effect.

In this letter we derive a new sum rule for the dynamical spin-density
structure function (SSF) which provides an independent
theoretical argument supporting this conjecture. It also predicts
that emissivities for weakly interacting particles should
increase somewhat slower than the lowest order rates at high
densities.

\section{The Spin-Density Structure Function}
In terms of the nucleon field operator in the nonrelativistic
limit, $\psi(x)$, the spin-density operator is given by
${\hbox{\boldmath $\sigma$}}(x)={1\over2}\psi^\dagger(x)
{\hbox{\boldmath $\tau$}}\psi(x)$ where
${\hbox{\boldmath $\tau$}}$ are the
Pauli matrices. In the following we denote the momentum,
coordinate, and spin operators for a single nucleon by ${\bf p}_i$,
${\bf r}_i$, and ${\hbox{\boldmath $\sigma$}}_i$, respectively, where
$i=1,\cdots,N_b$ runs
over $N_b$ nucleons. Then, for a normalization volume $V$, we
can define the Fourier transform
\begin{equation}
  {\hbox{\boldmath $\sigma$}}(t,{\bf k})={1\over V}\int d^3{\bf r}
  e^{-i{\bf k}\cdot{\bf r}}{\hbox{\boldmath $\sigma$}}(t,{\bf r})=
  {1\over V}\sum_{i=1}^{N_b}e^{-i{\bf k}\cdot{\bf r}_i}
  {\hbox{\boldmath $\sigma$}}_i\,.\label{sk}
\end{equation}
In terms of these operators and the baryon density $n_b$
the SSF is defined
as~\cite{R2,Janka}
\begin{equation}
  S_\sigma(\omega,{\bf k})={4\over3n_b}\int_{-\infty}^{+\infty}
  dte^{i\omega t}\left\langle{\hbox{\boldmath $\sigma$}}(t,{\bf k})
  \cdot{\hbox{\boldmath $\sigma$}}(0,-{\bf k})
  \right\rangle\,,\label{sdef}
\end{equation}
where $(\omega,{\bf k})$ is the four-momentum transfer to the
medium. The expectation value $\langle\cdots\rangle$ in
Eq.~(\ref{sdef}) is taken over a thermal ensemble.

The contribution of $S_\sigma$ to the neutrino scattering rate (per
final state density) from four momentum $(\omega_1,{\bf k}_1)$ to
$(\omega_2,{\bf k}_2)$ can be written as ${1\over4}G_{\rm
F}^2C_A^2n_b(3-\cos\,\theta)S_\sigma(\omega_1-\omega_2,{\bf k}_1-{\bf k}_2)$
with $G_{\rm F}$ the Fermi constant, $C_A$ the relevant
axial-vector charge, and $\theta$ the angle between ${\bf k}_1$
and ${\bf k}_2$. Similarly, the rate for pair production
would read
${1\over4}G_{\rm F}^2C_A^2n_b(3+\cos\,\theta)S_\sigma(-\omega_1-\omega_2,
-{\bf k}_1-{\bf k}_2)$~\cite{R2}. The axion emission rate per volume,
$Q_a$, is governed by the same structure function [in an
isotropic medium $S_\sigma(\omega,{\bf k})=S_\sigma(\omega,k)$ only
depends on $k=\vert{\bf k}\vert$]:
\begin{equation}
  Q_a={C_N^2n_b\over(4\pi)^2f_a^2}\int_0^\infty d\omega\,\omega^4
  S_\sigma(-\omega,\omega)\,.\label{Qa}
\end{equation}
Here, $f_a$ is the Peccei-Quinn scale and the numerical factor
$C_N$ depends on the specific axion model~\cite{R3}. Neutrino
opacities and axion emissivities are therfore mainly determined
by the SSF at thermal energies $\omega\simeq k\la T$.

Eq.~(\ref{sdef}) implies
\begin{equation}
  \int_{-\infty}^{+\infty}{d\omega\over2\pi}\,\omega S_\sigma(\omega,{\bf k})
  =-{4\over3n_b}\left\langle[H,{\hbox{\boldmath
  $\sigma$}}(0,{\bf k})]\cdot{\hbox{\boldmath $\sigma$}}(0,-{\bf k})
  \right\rangle\,,\label{sum1}
\end{equation}
where $H$ is the Hamiltonian of the system of interacting
nucleons for which we assume the following form:
\begin{equation}
  H=H_0+H_{\rm int}=\sum_{i=1}^{N_b}{{\bf p}_i^2\over2M}+{1\over2}
  \sum_{i\neq j}^{N_b}V({\bf r}_{ij},{\hbox{\boldmath $\sigma$}}_i,
  {\hbox{\boldmath $\sigma$}}_j)\,.\label{H}
\end{equation}
Here, ${\bf r}_{ij}={\bf r}_i-{\bf r}_j$, $M$ is the
free nucleon mass, and $V({\bf r}_{ij},{\hbox{\boldmath
$\sigma$}}_i,{\hbox{\boldmath $\sigma$}}_j)$ is the spin
dependent two nucleon interaction potential. For notational
simplicity we restrict ourselves to only one nucleon species for
the moment; the general case will be discussed further below.

For free nucleons one gets
$\int_{-\infty}^{+\infty}(d\omega/2\pi)\,\omega
S_\sigma(\omega,{\bf k})={\bf k}^2/2M$, in analogy to
the well known f sum rule for the dynamical density structure
function. In the latter case nucleon number conservation ensures
that the f sum rule even holds in the
presence of velocity independent interactions. In contrast, the
f sum for the SSF is modified in the presence of spin dependent
interactions since the nucleon spin is in general not conserved.

For one nucleon species the most general two nucleon interaction
potential is of the form ~\cite{ST}
\begin{equation}
  V({\bf r},{\hbox{\boldmath $\sigma$}}_1,{\hbox{\boldmath
  $\sigma$}}_2)=U(r)+U_S(r){\hbox{\boldmath $\sigma$}}_1
  \cdot{\hbox{\boldmath $\sigma$}}_2
  +U_T(r)\left(3{\hbox{\boldmath $\sigma$}}_1\cdot\hat{\bf r}\,
  {\hbox{\boldmath $\sigma$}}_2\cdot\hat{\bf r}-{\hbox{\boldmath
  $\sigma$}}_1\cdot{\hbox{\boldmath $\sigma$}}_2\right)
  \,,\label{Vint}
\end{equation}
where ${\bf r}={\bf r}_{12}$, $r=\vert{\bf r}\vert$, and $\hat{\bf
r}={\bf r}/r$. We denote the spin dependent terms by
$V^{S}_{ij}=U_S(r_{ij}){\hbox{\boldmath $\sigma$}}_i
\cdot{\hbox{\boldmath $\sigma$}}_j$ (the ``scalar force'') and
$V^{T}_{ij}=U_T(r_{ij})\left(3{\hbox{\boldmath
$\sigma$}}_i\cdot\hat{\bf r}_{ij}\,{\hbox{\boldmath $\sigma$}}_j\cdot
\hat{\bf r}_{ij}-{\hbox{\boldmath $\sigma$}}_i
\cdot{\hbox{\boldmath $\sigma$}}_j\right)$ (the ``tensor
force''). In order to
calculate the additional commutator in Eq.~(\ref{sum1})
from Eqs.~(\ref{sk}), (\ref{H}) and (\ref{Vint}) we make use of
the commutation relations
$\left[\sigma_i^a,\sigma_j^b\right]=
i\delta_{ij}\epsilon^{abc}\sigma_i^c$,
where $i,j=1,\cdots,N_b$ and $\epsilon^{abc}$ is the total
antisymmetric tensor in the spatial indices $a,b,c$.
After some algebra and using the symmetry
properties of the Hamiltonian the modified sum rule reads
\begin{equation}
  \int_{-\infty}^{+\infty}{d\omega\over2\pi}\,\omega
  S_\sigma(\omega,{\bf k})
  ={{\bf k}^2\over2M}\label{sum1a}
  -{4\over3N_b}\sum_{i\neq j}^{N_b}
  \left\langle V^{S}_{ij}+V^{T}_{ij}+\cos\,{\bf k}\cdot{\bf r}_{ij}
  \left({1\over2}V^{T}_{ij}-V^{S}_{ij}\right)
  \right\rangle\,.
\end{equation}
The kinetic nucleon recoil term is in general
negligible compared to the $V$-dependent terms which govern the
inelasticity of axial-vector interactions.
If $V({\bf r},{\hbox{\boldmath $\sigma$}}_1,{\hbox{\boldmath
$\sigma$}}_2)\ga-\alpha/r^s$ with $\alpha>0$ and $s<2$
the eigenvalues of $H$ are bounded from below and the r.h.s. of
Eq.~(\ref{sum1a}) is finite as long as $U_S(r)$ and $U_T(r)$
are integrable. This is the case for typical meson exchange
potentials with hard core repulsion~\cite{ST,BD,FM}.
Assuming the three terms in Eq.~(\ref{Vint}) to be of similar
size the r.h.s. of Eq.~(\ref{sum1a}) is roughly proportional to
the average interaction energy per nucleon $W$. At zero temperature
and for SN core densities and compositions,
$W\simeq30\,{\rm MeV}$ corresponding to an average binding energy
of about $10\,{\rm MeV}$ per nucleon. For $T\ga10\,{\rm MeV}$
nucleons are bound more weakly and $W$ should be considerably
smaller. We can therefore write
\begin{equation}
  \int_{-\infty}^{+\infty}{d\omega\over2\pi}\,\omega
  S_\sigma(\omega,{\bf k})\simeq4W
  \la100\,{\rm MeV}\,,\label{Vmax}
\end{equation}
where the inequality is a conservative bound reflecting our poor
knowledge about the equation of state for hot nuclear matter.
Since it involves bound state energies, Eq.~(\ref{Vmax}) is a
nonperturbative result and will play an
important role for the high density behavior of weak interaction
rates below.

The dependence on the momentum transfer ${\bf k}$ in
Eq.~(\ref{sum1a}) is expected to be only modest.
In fact, for $k\la T\la50\,{\rm MeV}$, we
have $\vert{\bf
k}\cdot{\bf r}\vert\ll1$ within the range of the potential
$r_s\simeq1/m_\pi$ which is determined by the pion mass
$m_\pi\simeq140\,{\rm MeV}$. We can thus go to the long
wavelength limit~\cite{Iwamoto1,R2,Sawyer,Keil,Janka},
${\bf k}\rightarrow0$, using
$S_\sigma(\omega)\equiv S_\sigma(\omega,{\bf k}\rightarrow0)$.
Eq.~(\ref{sum1a}) then simplifies to
\begin{equation}
  \int_{-\infty}^{+\infty}{d\omega\over2\pi}\,\omega S_\sigma(\omega)
  =-{4\over N_b}\langle H_T\rangle
  \,,\label{sum1b}
\end{equation}
where $H_T={1\over2}\sum_{i\neq j}^{N_b}V^{T}_{ij}$.

First, note that the scalar force does not contribute to
Eq.~(\ref{sum1b}) because it
conserves the total nucleon spin ${\hbox{\boldmath
$\sigma$}}(0,{\bf k}\rightarrow0)$
[see Eq.~(\ref{sk})]. Below nuclear densities the
nucleon-nucleon ($NN$) interaction is dominated by one-pion exchange
(OPE) leading to a tensor force.
This contribution induces a spin orbit coupling and does therefore not
conserve the total nucleon spin. Thus only the tensor force
contributes to Eq.~(\ref{sum1}) in the long
wavelength limit. This agrees with the lowest order
bremsstrahlung calculation for ${\bf k}\rightarrow0$~\cite{FM}.
Finally, note that the r.h.s. of Eq.~(\ref{sum1b}) is
positive as it should be since the interaction induced
correlations reduce $\langle H_T\rangle$ below the value for
free nucleons, $\langle H_T\rangle=0$.

An additional sum rule~\cite{R2,Keil,Janka} can be
obtained by integrating Eq.~(\ref{sdef}) and using
Eq.~(\ref{sk}) in the long wavelength limit:
\begin{equation}
  \int_{-\infty}^{+\infty}{d\omega\over2\pi}\,S_\sigma(\omega)
  =1+{4\over3N_b}\left\langle\sum_{i\neq j}^{N_b}{\hbox{\boldmath
  $\sigma$}}_i\cdot{\hbox{\boldmath $\sigma$}}_j\right\rangle
  \,.\label{sum2}
\end{equation}
Note that for free nucleons Eqs.~(\ref{sum1b}) and (\ref{sum2})
yield $S_\sigma(\omega)=2\pi\delta(\omega)$, whence only elastic
scattering on the medium is possible in the absence of $NN$
interactions.

\section{Dilute Medium Limit}
At low densities, i.e. for
large average inter-nucleon spacing, the interaction energy $W$
in Eq.~(\ref{Vmax}) is much smaller than the kinetic terms from
the free Hamiltonian. In case of the long wavelength limit,
Eq.~(\ref{sum1b}), we can therefore treat $H_T$ as a small
perturbation and write to lowest non-trivial order in $H_T$:
\begin{equation}
  \langle H_T\rangle={2\over Z}\sum_n\exp\left(-E_n^0/T\right)
  {\rm Re}\left[_1\!\left\langle n\right\vert H_T\left\vert
  n\right\rangle_0\right]\,.\label{H31}
\end{equation}
Here, $E_n^0$, $\left\vert n\right\rangle_0$ are the eigenvalues and
eigenstates of the free Hamiltonian $H_0$, respectively,
$\left\vert n\right\rangle_1$ are the eigenvectors of $H_0+H_T$
to first order
in $H_T$, and $Z=\sum_n\exp\left(-E_n^0/T\right)$ is the normalization
factor. Assuming nondegenerate eigenstates for simplicity and
applying standard first order perturbation theory for
$\left\vert n\right\rangle_1$ we can express everything in terms
of zeroth order quantities.
Dropping the index $0$ from now on, Eq.~(\ref{H31}) reduces to
the negative definite expression
\begin{equation}
  \langle H_T\rangle={1\over Z}\sum_{n\neq m}
  {e^{-E_n/T}-e^{-E_m/T}\over
  E_n-E_m}\left\vert(H_T)_{mn}\right\vert^2\,,\label{H32}
\end{equation}
where $(H_T)_{mn}=_0\!\left\langle m\right\vert H_T\left\vert
n\right\rangle_0$. This matrix element is expected to vary in
$E_m-E_n$ over a scale $\ga3m_\pi^2/M\simeq50\,{\rm MeV}$ where
$m_\pi$ is a typical momentum scale in the $NN$ interaction
potential. Therefore, for
$T\la50\,{\rm MeV}$ the thermal factor in Eq.~(\ref{H32})
can be approximated by $\delta(E_m-E_n)$. Converting the sum
over $m$ into an integral over $E_m$ Fermi's golden rule finally
gives $W\simeq-\langle H_T\rangle/N_b=\Gamma_\sigma/(2\pi)$. Here,
$\Gamma_\sigma$ is the average perturbative $NN$ scattering
rate mediated by $H_T$ which is a measure for
the spin fluctuation rate. The spins fluctuate on a time scale
given by the inverse energy scale of the tensor force
which causes the spin fluctuations. At high densities we use
$\Gamma_\sigma\equiv-2\pi\langle H_T\rangle/N_b$ as an effective
spin flip rate.

\section{Saturation of Spin Fluctuation Rates}
We now use the sum rules Eqs.~(\ref{sum1b}) and (\ref{sum2}) to
determine the qualitative form of $S_\sigma(\omega,{\bf k})$ in the long
wavelength limit. To this end let us introduce the dimensionless
quantity $\tilde{S}_\sigma(x)\equiv TS_\sigma(xT)$ with
$x=\omega/T$ as in
Ref.~\cite{Keil}. Due to the principle of detailed balance,
$S_\sigma(\omega,{\bf k})=S_\sigma(-\omega,-{\bf
k})e^{\omega/T}$, it is sufficient
to specify $\tilde{S}_\sigma(x)$ for $x>0$ only.

Introducing the dimensionless effective spin flip rate
$\gamma_\sigma=\Gamma_\sigma/T$, we can write the sum rule
Eq.~(\ref{sum1b}) as
\begin{equation}
  \int_0^{+\infty}{dx\over2\pi}x\tilde{S}_\sigma(x)\left(1-e^{-x}\right)
  ={2\gamma_\sigma\over\pi}\simeq{4W\over T}\,,\label{sum1d}
\end{equation}
where in a newly born neutron star
$\gamma_\sigma$ does not increase beyond a few.

Furthermore, since in a hot SN core the thermal energies are
expected to be considerably higher than the interaction energy
$W$, within a first approximation we can neglect spin
correlations in the second sum rule Eq.~(\ref{sum2}) and write
\begin{equation}
  \int_0^{+\infty}{dx\over2\pi}\tilde{S}_\sigma(x)\left(1+e^{-x}\right)
  \simeq1\,.\label{sum2a}
\end{equation}

For the following discussion we consider the general case of an
ensemble of neutrons and protons with fractional number
densities $Y_n$ and $Y_p$. Introducing the isospin operators
${\hbox{\boldmath $\tau$}}_i$ for nucleon $i$,
${\hbox{\boldmath $\sigma$}}_i$
in the definition of $S_\sigma$ [see Eqs.~(\ref{sdef}) and (\ref{sk})]
has to be multiplied by $\left[1+(\tau_i)_3\right]C_{A,p}/2
+\left[1-(\tau_i)_3\right]C_{A,n}/2$. Here, $C_{A,p}$
and $C_{A,n}$ are the relevant proton and neutron axial-vector
charges. Moreover, there will be additional terms
proportional to ${\hbox{\boldmath $\tau$}}_i
\cdot{\hbox{\boldmath $\tau$}}_j$ in the interaction potential
Eq.~(\ref{Vint}). However, this leaves our discussion
qualitatively unchanged since the additional isospin operators appearing
under the expectation values only lead to additional factors of order
unity. If correlations among different nucleons are absent the
r.h.s. of the sum rules Eqs.~(\ref{sum1d}) and (\ref{sum2a}) get
multiplied by
$(Y_pC_{A,p}^2+Y_nC_{A,n}^2)/(C_{A,p}^2+C_{A,n}^2)$.

Parametrizing the high $\omega$ behavior of $S_\sigma$ by
$S_\sigma(\omega)\propto\omega^{-n}$, classical collisions would
lead to $n=2$. On the quantum mechanical level the deviation of
$S_\sigma(\omega)$ from $2\pi\delta(\omega)$ is to lowest order
in the strong interactions given by nucleon bremsstrahlung.
Using a dipole like OPE potential
without a hard core cutoff yields $n=5/2$ and $n=3/2$ in the case of
one and two nucleon species,
respectively~\cite{R2,FM,Iwamoto2,Brinkmann}. The non-existence
of the f sum Eq.~(\ref{sum1d}) in the latter
case stems from the unphysical $r^{-3}$ divergence
of this potential at $r=0$. Except for s waves this
divergence indeed leads to an infinite $\langle H_T\rangle$. If one
regularizes the potential by a hard core
repulsion f sum integrability is restored.

This motivates the following representative ansatz:
\begin{equation}
  \tilde{S}_\sigma(x)={a\over x^{5/2}+b}\quad\hbox{for}\;x>0
  \,,\label{St}
\end{equation}
where $a$ and $b$ are positive constants. The
sum rule Eq.~(\ref{sum1d}) is sensitive to the high energy
behavior and therefore mainly to $a$. In contrast, the sum rule
Eq.~(\ref{sum2a}) probes the ``infrared'' regime which is
sensitive to $b$. Eq.~(\ref{St}) is of the form expected from
nucleon bremsstrahlung where $b$ accounts for the LPM effect.

We can now pick
a number for $a$, determine the corresponding value of
$b$ numerically from Eq.~(\ref{sum2a}) and compute
the f sum Eq.~(\ref{sum1d}).
The result is plotted in Fig.~1 as a function of $a$.
Most importantly, from the expected density and temperature
dependence of $W$ we expect the f sum to increase monotonically
towards the SN core before saturating at a value of order unity.
As a consequence, the thermally averaged axial-vector neutrino
scattering cross section $\langle\sigma_A\rangle$ which
dominates the neutrino opacity should roughly scale
as $T^2$ being density independent as naively expected (see
Fig.~1). Furthermore, the axion emission rate from
Eq.~(\ref{Qa}) approximately scales as $n_b\Gamma_\sigma T^3$.
The lowest order axion emissivities should therefore be
multiplied by $\Gamma_\sigma/\Gamma_\sigma^\prime$ whenever this
ratio is smaller than 1. Here, $\Gamma_\sigma^\prime$ is the
lowest order spin flip rate extrapolated from the dilute medium
limit. For example, $\Gamma_\sigma^\prime\simeq32\,{\rm
MeV}\rho_{14}T_{10}^{1/2}$ for the standard OPE
calculations~\cite{FM,Iwamoto2,Brinkmann}, where $\rho_{14}$ is
the mass density in $10^{14}\,{\rm gcm}^{-3}$ and
$T_{10}=T/10\,{\rm MeV}$. A turn over in
$\langle\sigma_A\rangle/T^2$ and $Q_a/(n_bT^3)$ typically only
occurs at $\gamma_\sigma\ga10$ and is the less
pronounced the stronger $S_\sigma(\omega)$ falls off at large
$\omega$. The absence of a decrease of these quantities at high
density is therefore rather independent of
uncertainties in the exact saturation value for
$\gamma_\sigma$.

\section{Summary}
Neutrino opacities and axion
emissivities are governed mainly by the SSF. We have derived a
new sum rule for the SSF
which corresponds to the f sum rule for the density structure
function but depends on the nucleon spin flip
interactions. Our treatment so far
assumes absence of possible pion and kaon condensates. Employing
an infrared regularized bremsstrahlung spectrum for the
functional form of the SSF we have shown that the effective spin
fluctuation rate $\Gamma_\sigma$ must saturate somewhere
below $\simeq150\,{\rm MeV}$ which is within factors of a
few of SN core temperatures.
Neutrino scattering cross sections should therefore
exhibit the naive $T^2$ scaling whereas axion emissivities
should increase somewhat slower than the lowest order rates at
high densities. There is no turnover of weak
interaction rates towards the SN core. These results
have an important impact on SN cooling simulations and their
application to the derivation of axion mass bounds. They are also
relevant for the rates for URCA processes and emission of right
handed neutrinos.

\section*{Acknowledgments}
I gratefully acknowledge Georg Raffelt for an extensive e-mail
correspondence on many aspects of this research and for providing me
with early versions of Ref.~\cite{Janka}. I also thank
Thomas Janka for discussions of various astrophysical aspects.
This work was supported by the DoE, NSF and NASA at the
University of Chicago, by the DoE and by NASA through grant
NAG5-2788 at Fermilab, and by the Alexander-von-Humboldt
Foundation. Furthermore, I wish to thank the Aspen Center for
Physics where part of this research has been conducted for
hospitality and financial support.

\newpage

\section*{Figure Captions}
\bigskip

\noindent{\bf Fig.~1. }
The f sum Eq.~(\ref{sum1d}) characterized by
$\gamma_\sigma$ as a function of the parameter $a$
(solid line). Also shown in arbitrary units
are the axion emission rate per baryon $Q_a/n_b$ (dashed line) and the
thermal axial-vector neutrino scattering cross section
$\langle\sigma_A\rangle$ (dotted line) normalized to a fixed temperature.
The physical range is where the f sum is smaller than a few.

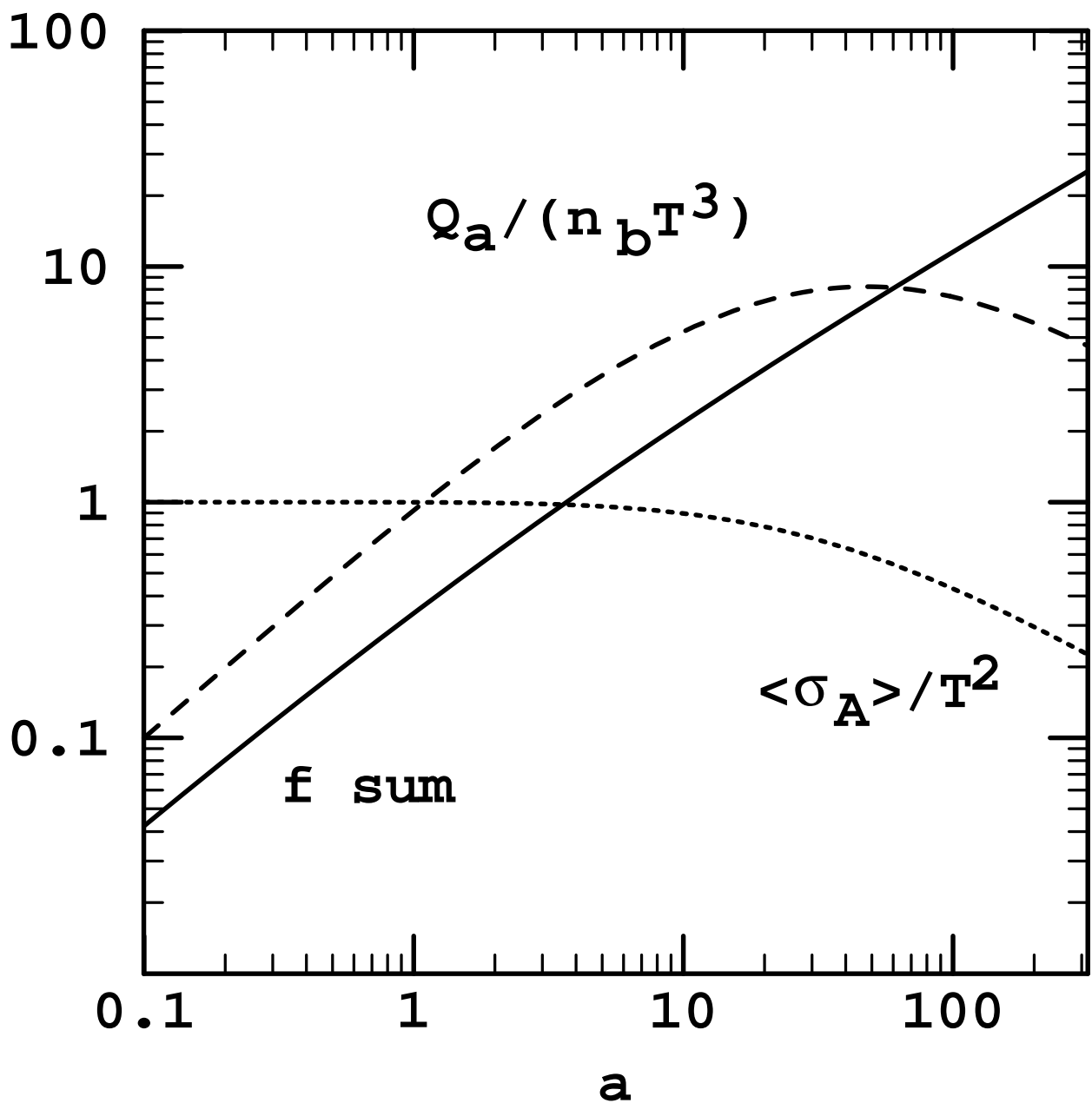
\end{document}